\documentclass[aps,prc,preprint,showpacs,superscriptaddress,amsmath,amssymb,floatfix,nofootinbib]{revtex4}

\usepackage[dvips]{graphicx,color}
\usepackage{dcolumn}
\usepackage{bm}
\usepackage{here}

\def\pmb#1{\setbox0=\hbox{#1}%
  \kern-.025em\copy0\kern-\wd0 
  \kern.05em\copy0\kern-\wd0
  \kern-.025em\raise.0433em\box0 }

\def\lambdabar{\protect\@lambdabar}
\def\@lambdabar{%
\relax
\bgroup
\def\@tempa{\hbox{\raise.73\ht0
\hbox to0pt{\kern.25\wd0\vrule width.5\wd0
height.1pt depth.1pt\hss}\box0}}%
\mathchoice{\setbox0\hbox{$\displaystyle\lambda$}\@tempa}%
{\setbox0\hbox{$\textstyle\lambda$}\@tempa}%
{\setbox0\hbox{$\scriptstyle\lambda$}\@tempa}%
{\setbox0\hbox{$\scriptscriptstyle\lambda$}\@tempa}%
\egroup
}

\begin{document}

\preprint{J-PARC-TH-0214}

\title{\boldmath
Medium effects on $\Xi^-$ production in the nuclear ($K^-$,~$K^+$) reaction
}

\author{Toru~Harada}%
\email{harada@osakac.ac.jp}
\affiliation{%
Research Center for Physics and Mathematics,
Osaka Electro-Communication University, Neyagawa, Osaka, 572-8530, Japan
}
\affiliation{%
J-PARC Branch, KEK Theory Center, Institute of Particle and Nuclear Studies,
High Energy Accelerator Research Organization (KEK),
203-1, Shirakata, Tokai, Ibaraki, 319-1106, Japan
}
\author{Yoshiharu~Hirabayashi}%
\affiliation{%
Information Initiative Center, 
Hokkaido University, Sapporo, 060-0811, Japan
}

\date{\today}

\begin{abstract}
We study theoretically medium effects on $\Xi^-$ production in the ($K^-$,~$K^+$) reaction, 
using the optimal Fermi-averaging procedure which describes the Fermi motion of
a nucleon  on the on-energy-shell $K^-p\to K^+\Xi^-$ reaction condition in nuclei.
The result shows the strong energy and angular dependence of the in-medium 
$K^-p\to K^+\Xi^-$ cross section,
which affects significantly the shape and magnitude of the production 
spectrum for $\Xi^-$ hypernuclear states in the ($K^-$,~$K^+$) reaction on a nuclear target.
The application to the $\Xi^-$ quasi-free production via the ($K^-$,~$K^+$) reaction on a $^{12}$C target 
is also discussed in a Fermi gas model. 
\end{abstract}

\maketitle


\section{Introduction}
\label{Intro}

It is important to understand properties of $\Xi$ hypernuclei 
whose states are regarded as the opening of the $S=-2$ world in nuclear physics.
This is a significant step to extend studies of 
multi-strangeness systems and also strange neutron stars in astrophysics \cite{Chatterjee16}.
Recently, Nakazawa et al.~\cite{Nakazawa15} reported the first evidence of a bound state 
of the $\Xi^-$-$^{14}$N system which was identified by the ``KISO'' event 
in the KEK-E373 experiment.
This result supported that the $\Xi$-nucleus potential has a weak attraction of $V_\Xi\simeq$ 14 MeV
in the Wood-Saxon (WS) potential, as suggested by previous analyses \cite{Tadokoro95,Fukuda98,Khaustov00}.
However, there still remains an uncertainty about the nature of the $S=-2$ dynamics 
caused by the $\Xi N$ interaction and also $\Xi N$--$\Lambda\Lambda$ coupling in nuclei 
due to the limit to the available data. 
More experimental information is needed for the understanding of $\Xi^-$ hypernuclei.  

A pioneer study of $\Xi$ hypernuclei by Dover and Gal \cite{Dover83} 
indecated that a $\Xi$-nucleus potential has a well depth of $V_\Xi=$ $24 \pm 4$ MeV 
on the analysis of old emulsion data.
Khaustov et al.~\cite{Khaustov00} discussed a missing mass spectrum near the $\Xi$ threshold 
in the $^{12}$C($K^-$,~$K^+$) reaction at 1.8 GeV/$c$ at the BNL-E885 experiment. 
Their analysis showed $V_\Xi\simeq$ 14 MeV for the $\Xi$-nucleus potential, 
whereas the resolution was not sufficient to resolve $\Xi^-$ hypernuclear states.
Recently, Nagae et al.~\cite{Nagae18} have performed an accurate 
observation of the $\Xi^-$ production spectrum in the $^{12}$C($K^-$,~$K^+$) reaction 
at 1.8 GeV/$c$ in the J-PARC E05 experiment, and their analysis is now ongoing.

On the other hand, the authors \cite{Harada05,Harada06,Harada18} discussed 
the $\Sigma^-$-nucleus potential in the quasi-free (QF) spectra of the ($\pi^-$,~$K^+$) 
reactions on the nuclear targets, 
using the optimal Fermi-averaging procedure \cite{Harada04}. 
This procedure describes the Fermi motion of a nucleon 
on the on-energy-shell $\pi^-p\to K^+\Sigma^-$ reaction condition in nuclei, 
and it generates the energy dependence of the in-medium $\pi^-p \to K^+\Sigma^-$ 
reaction amplitude. 
This energy dependence due to the Fermi motion leads to a successful explanation 
of the ($\pi^-$,~$K^+$) spectra in the DWIA analyses on $^{28}$Si \cite{Harada05}, 
$^{209}$Bi \cite{Harada06}, and $^{6}$Li \cite{Harada18} targets; 
it is possible to extract properties of the $\Sigma$-nucleus potential appropriately 
from the experimental data in nuclear ($\pi^-$,~$K^+$) reactions. 
Therefore, it is worthwhile studying the medium effects on $\Xi^-$ production 
in the nuclear ($K^-$,~$K^+$) reactions, employing the optimal Fermi-averaging procedure.

Maekawa et al.~\cite{Maekawa07} investigated the $\Xi^-$ QF spectra in 
the ($K^-$,~$K^+$) reaction on $^{12}$C in the similar 
calculations considering the local momentum for the nucleon and $\Xi$.
Hashimoto et al.~\cite{Hashimoto08} also discussed 
the $\Xi^-$ QF spectra in the ($K^-$,$K^+$) reactions on $^{12}$C
in the semiclassical distorted wave (SCDW) method. 
Recently, Kohno \cite{Kohno19} has reexamined the $\Xi^-$ QF spectra 
in the ($K^-$,$K^+$) reactions on $^9$Be and $^{12}$C targets 
in the SCDW method, using the $\Xi$-nucleus potential derived from 
the next-to-leading order (NLO) in chiral effective field theory (ChEFT).
However, it seems that the calculated $\Xi^-$ QF spectra are 
insufficient to reproduce the data quantitatively. 

In this paper, 
we study theoretically medium effects on $\Xi^-$ production in the nuclear ($K^-$,$K^+$) reaction. 
We evaluate the in-medium cross sections of the $K^-p\to K^+\Xi^-$ reaction, 
using the optimal Fermi-averaging procedure \cite{Harada04}, 
and we discuss the energy and angular dependence of the $\Xi^-$ production spectrum 
in the ($K^-$,~$K^+$) reaction on a nuclear target.
We also apply this procedure to the $\Xi^-$ QF production spectrum in the ($K^-$,~$K^+$) reaction 
on a $^{12}$C target at $p_{K^-}=$ 1.8 GeV/$c$ in a Fermi gas model.

\section{Calculations}

\subsection{Distorted-wave impulse approximation}

Let us consider a calculation procedure of the hypernuclear production
 for the nuclear ($K^-$,~$K^+$) reaction in the laboratory frame. 
The inclusive double differential cross sections within 
the distorted-wave impulse approximation (DWIA)
\cite{Hufner74,Auerbach83} are given by 
(in units $\hbar=c=1$)
\begin{eqnarray}
   {{d}^2{\sigma} \over {d}E_{K^+}{d}\Omega_{K^+} }
&=& \beta
    {1 \over {[J_A]}}
    \sum_{m_A}\sum_{B,m_B} \vert\langle {\Psi}_B \vert
    \hat{F} \vert {\Psi}_A \rangle\vert^{2} \nonumber\\
&& \times \delta (\omega-E_{B}+E_{A}), 
\label{eqn:e1}
\end{eqnarray}
where $[J]=2J+1$, $\Psi_B$ and $\Psi_A$ ($E_{B}$ and $E_{A}$)
are wavefunctions (energies) of hypernuclear final states 
and an initial state of the target nucleus, respectively.
The laboratory energy and momentum transfer is
\begin{eqnarray}
\omega= E_{K^-}-E_{K^+}, \quad {\bm q}={\bm p}_{K^-} - {\bm p}_{K^+},
\label{eqn:e1a}
\end{eqnarray}
where $E_{K^-}$ and ${\bm p}_{K^-}$ ($E_{K^+}$ and ${\bm p}_{K^+}$) denote 
an energy and a momentum of the incoming $K^-$ (the outgoing $K^+$),
respectively. 
The energy transfer may also be expressed as
\begin{eqnarray}
\omega = E_B -E_A = m_\Xi - m_N -B_\Xi -\varepsilon_N + T_{\rm recoil},
\label{eqn:e1b}
\end{eqnarray}
where $B_\Xi$ is a $\Xi^-$ binding energy, $\varepsilon_N$ is a single-particle energy 
of a nucleon-hole state, and $T_{\rm recoil}$ is a recoil energy to the final state.
The kinematical factor $\beta$ \cite{Tadokoro95} 
arising from a translation from a two-body meson-nucleon laboratory system to a meson-nucleus 
laboratory system \cite{Dover83} is given by 
\begin{equation}
 \beta=
 \biggl(1+ {E^{(0)}_{K^+} \over E^{(0)}_{B}}
        {{p^{(0)}_{K^+} - p^{(0)}_{K^-} \cos\theta_{\rm lab}}
        \over p^{(0)}_{K^+}} \biggr)
        {p_{K^+} E_{K^+} \over p^{(0)}_{K^+} E^{(0)}_{K^+}},
\label{eqn:e2}
\end{equation}
where 
$p^{(0)}_{K^-}$ and $p^{(0)}_{K^+}$ ($E^{(0)}_{K^+}$ and $E^{(0)}_{B}$) are 
laboratory momenta of $K^-$ and $K^+$ (laboratory energies of $K^+$ and $\Xi^-$)
in the two-body $K^- p\to K^+\Xi^-$ reaction, respectively.
Here we considered only the non-spin-flip reaction because 
the spin-flip contribution seems to be not so large in the ($K^-$,~$K^+$) reaction.
Thus an external operator $\hat{F}$ for the associated production 
$K^-p \to K^- \Xi^-$ reactions is given by
\begin{eqnarray}
\hat{F}
 &=& 
\int d{\bm r} \> \chi_{{\bm p}_{K^+}}^{(-) \ast}({\bm r})
\chi_{{\bm p}_{K^-}}^{(+)}({\bm r}) \nonumber\\
 &&  \times \sum_{j=1}^A \overline{f}_{K^-p\to K^+\Xi^-}
\delta ({\bm r}-{\bm r}_{j}){\hat O}_j, 
\label{eqn:e3}
\end{eqnarray}
with zero-range interaction for the $K^-p\to K^+\Xi^-$ transitions.
$\chi_{K^+}^{(-) \ast}$ and $\chi_{K^-}^{(+)}$ are distorted waves of 
incoming $K^-$ and outgoing $K^+$, respectively.
${\hat O}_j$ is a baryon operator changing 
$j$th nucleon into a $\Xi^-$ hyperon in the nucleus, 
and ${\bm r}$ is the relative coordinate between the mesons and 
the center-of-mass (c.m.)~of the nucleus; $\overline{f}_{K^- p\to K^+\Xi^-}$
is the in-medium $K^-p\to K^+\Xi^-$ amplitude 
on the laboratory frame.

In the DWIA, a Fermi-averaged amplitude is often used for $\overline{f}_{K^- p\to K^+\Xi^-}$ 
so as to take into account a Fermi motion in the nuclear medium \cite{Auerbach83}.
However, it is should be noticed that the energy dependence of 
$\overline{f}_{K^-p \to K^+\Xi^-}$ may play an important role in 
describing the QF spectrum within a wide energy range 
($\sim$ a few hundred MeV) as well as the angular dependence.
The authors \cite{Harada05,Harada06,Harada18} emphasized
the importance of the energy dependence of the Fermi-averaged amplitude of 
${\overline{f}}_{\pi^-p \to K^+\Sigma^-}$ in the nuclear ($\pi^-$,~$K^+$) reactions 
in order to extract properties of the $\Sigma$-nucleus potential 
from the experimental data, using the optimal Fermi-averaging procedure \cite{Harada04}.
Therefore, we are interested in the 
energy and angular dependence of $\overline{f}_{K^- p\to K^+\Xi^-}$ 
which is generated by the optimal Fermi-averaging procedure 
in the nuclear ($K^-$,~$K^+$) reaction.

\subsection{Optimal Fermi averaging}

According to Ref.~\cite{Harada04}, we consider the optimal Fermi averaging 
for the $K^-p \to K^+ \Xi^-$ reaction in a nucleus. 
To see clearly the medium effects of the $K^-p \to K^+ \Xi^-$ processes in 
the nuclear ($K^-$,~$K^+$) reaction in the framework of the DWIA, 
we introduce an ``optimal'' cross section for the $K^-p \to K^+ \Xi^-$ processes
in the nucleus, which can be given as 
\begin{eqnarray}
\Bigl( \displaystyle{d \sigma \over d \Omega} \Bigr)^{\rm opt}_{\theta_{\rm lab}}
&\equiv& 
\beta |\overline{f}_{K^-p \to K^+\Xi^-}|^2 \nonumber\\
& =&
{ p_{K^+} E_{K^+} \over (2\pi)^2 v_{K^-}}|t^{\rm opt}_{\bar{K}N,K\Xi}(p_{K^-}; 
\omega,{\bm q})|^2,
\label{eqn:e5}
\end{eqnarray}
where $v_{K^-}=p_{K^-}/E_{K^-}$.
The optimal Fermi-averaged $\bar{K}N\to K \Xi$ $t$ matrix, 
${t}^{\rm opt}_{\bar{K}N,K\Xi}(p_{\bar{K}}; \omega,{\bm q})$, is defined by 
\begin{eqnarray}
&&{t}^{\rm opt}_{\bar{K}N,K\Xi}(p_{\bar{K}}; \omega,{\bm q}) \nonumber\\
&&=
{\int^{\pi}_{0} \sin{\theta_N} d \theta_N \int_{0}^{\infty} 
dp_{N} p_N^2 \rho{(p_N)}{t}_{\bar{K}N,K \Xi}(E_{2};{\bm p}_{\bar{K}},{\bm p}_N)
\over
\int^{\pi}_{0} \sin{\theta_N} d \theta_N \int_{0}^{\infty} 
dp_{N} p_N^2 \rho{(p_N)}
  }\Biggl|_{{\bm p}_N={\bm p}^*_N}, 
\label{eqn:e6}
\end{eqnarray}
where $E_N$ and ${\bm p}_N$ are an energy and a momentum of a proton
in the target nucleus, respectively; 
$\cos{\theta_N}= \hat{\bm p}_{\bar{K}}\cdot\hat{\bm p}_N$, 
$E_{2}=E_{K^-}+E_{N}$ is a total energy of the $K^- N$ system, 
and $\rho(p_N)$ is a Fermi-momentum distribution of the proton in the target nucleus. 
The momentum ${\bm p}_N^*$ in Eq.~(\ref{eqn:e6}) is a solution which satisfies 
the on-energy-shell equation for a struck proton in the nuclear systems, 
\begin{eqnarray}
\sqrt{({\bm p}_N^*+{\bm q})^2+m_\Xi^2}-\sqrt{({\bm p}_N^*)^2+m_N^2}=\omega,
\label{eqn:e7}
\end{eqnarray}
where $m_\Xi$ and $m_N$ are masses of the $\Xi^-$ and the proton, respectively. 
This procedure keeps the on-energy-shell 
$K^- p \to K^+ \Xi^-$ processes in the nucleus \cite{Gurvitz86}, 
thus it guarantees to take ``optimal'' values for $\overline{f}_{K^-p \to K^+\Xi^-}$
within a factorized form, e.~g., see Eq.~(\ref{eqn:e8}). 
Note that binding effects for the nucleon and $\Xi$ in the nucleus 
are considered automatically 
when we input experimental values for the binding energies 
of the nuclear and hypernuclear states in Eq.~(\ref{eqn:e1b}). 
Here we neglected the energy dependence of a phase for the
$K^-p \to K^+\Xi^-$ $t$ matrix, and replaced 
${t}_{\bar{K}N,K\Xi}(E_{2};{\bm p}_{\bar{K}},{\bm p}_N)$ in the laboratory frame
by its absolute value 
$|{t}_{\bar{K}N,K\Xi}(E_{2};{\bm p}_{\bar{K}},{\bm p}_N)|$  which is obtained 
from the corresponding one in the c.m.~frame; 
${t}_{\bar{K}N,K\Xi}(E_{2};{\bm p}_{\bar{K}},{\bm p}_N)= \eta\, t_{\rm c.m.}(E_{\rm c.m.})$,
where $\eta$ is the M\"oller factor.
Such an assumption has been confirmed to be appropriate in the case of
the $\pi^+ n \to K^+ \Lambda$ process 
in the nuclear ($\pi^+$,~$K^+$) reactions \cite{Harada04}.

\subsection{${\bm K}^- {\bm p}\to {\bm K}^+ {\bm \Xi}^-$ reaction}

\begin{figure}[tb]
\begin{center}
  \includegraphics[width=0.8\linewidth]{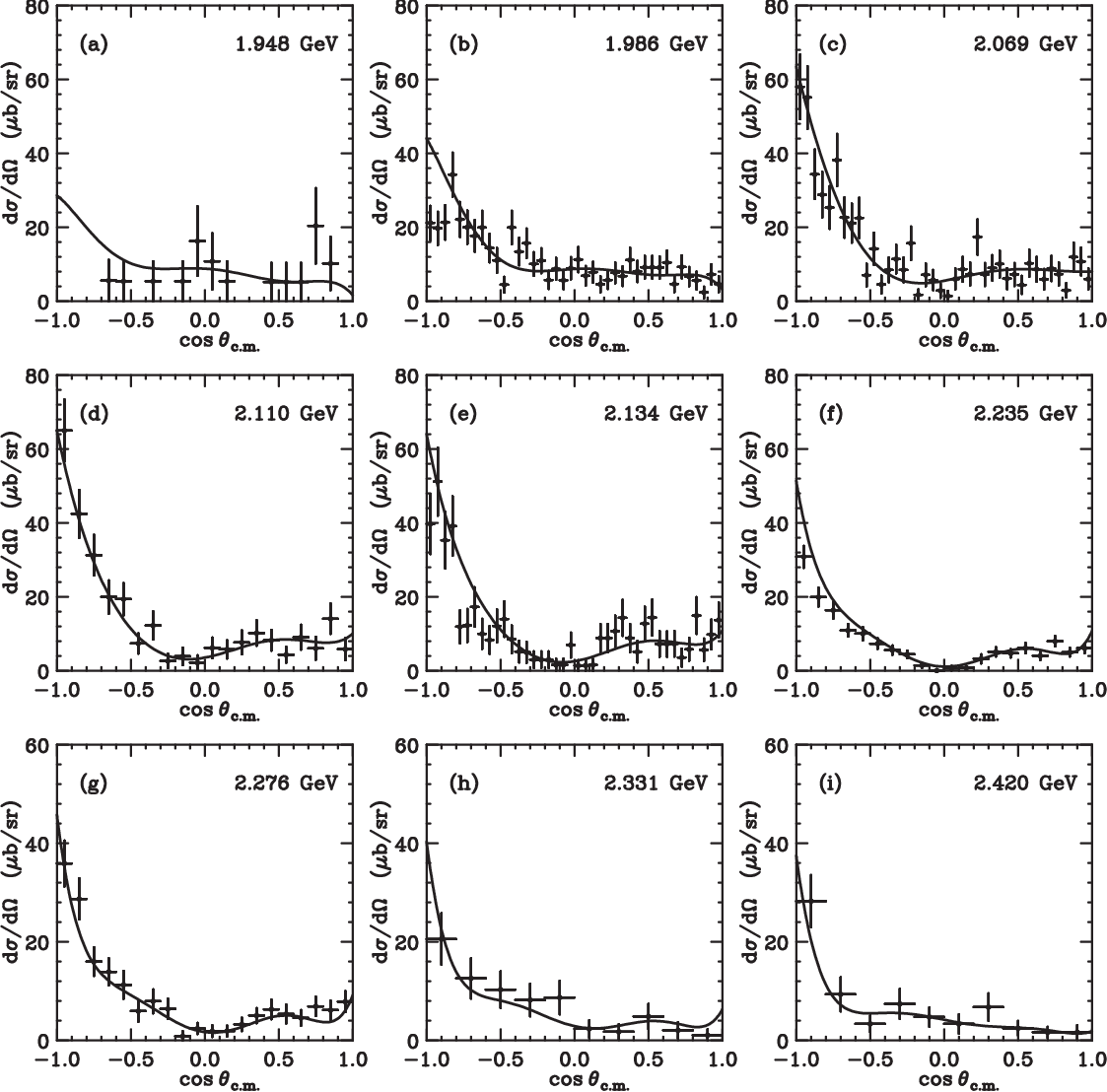}
\end{center}
\caption{\label{fig:1}
  Angular distributions of the differential cross section for 
  the $K^- p \to K^+ \Xi^-$ reaction in the c.m.~frame at 
  $E_{\rm c.m.}$ = 1.948, 1.966, 2.069, 2.110, 2.134, 2.235, 2.276, 2.331, 
and 2.420 GeV.
Solid curves denote the calculated values to make a fit to the experimental data.
The data are taken from Refs.~\cite{Trower68,Burgun68,Dauber69,Trippe67,London66}, 
following to the compilation in Ref.~\cite{Sharov11}.
}
\end{figure}

\begin{figure}[tb]
  \begin{center}
  \includegraphics[width=1.00\linewidth]{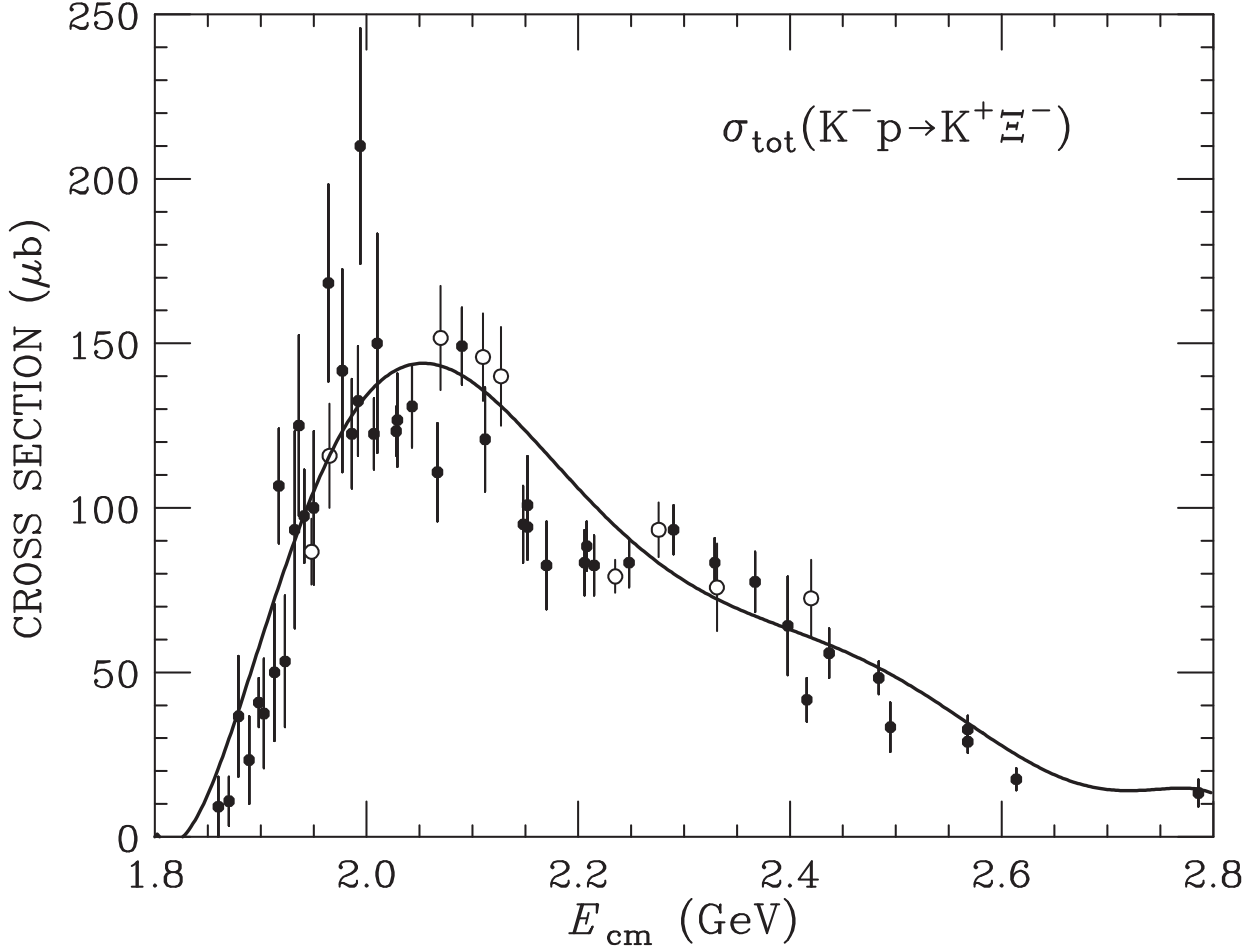}
  \end{center}
  \caption{\label{fig:2}
  Total cross section $\sigma_{\rm tot}$ for the $K^-p \to K^+\Xi^-$ reaction as a function 
of the energy $E_{\rm c.m.}$.
The data are taken from the compilation of 
Flaminio et al. \cite{Flaminio79}. 
Open and solid circles denote the data for $\sigma_{\rm tot}$ with and without 
the measurements of $d\sigma/d\Omega$ shown in Fig.~\ref{fig:1}, respectively. 
  } 
\end{figure}

In the optimal Fermi-averaging procedure, we need to prepare 
the elementary $K^- p \to K^+ \Xi^-$ $t$ matrices (amplitudes) which fully 
reproduce the experimental data of differential cross sections in free space.
Recently, several authors \cite{Sharov11,Jackson15} investigated 
the $K^- p \to K^+ \Xi^-$ reaction amplitudes,  
of which the nature is caused by $Y^*$ resonances as intermediate states
via the $K^-p \to K^+ \Xi^-$ processes,  
by the help of theoretical models due to poor quality in the available 
data \cite{Sharov11,Jackson15,Flaminio79}.
However, we use the angular distributions of the $K^- p \to K^+ \Xi^-$ reaction 
for the sake of simplicity. They are parametrized according to 
\begin{eqnarray}
\Bigl( {d \sigma \over d \Omega} \Bigr)^{\rm elem}_{\rm c.m.}
& =&
{\omega_{f}  \omega_{i} p_{f} \over (2\pi)^2 p_{i}}
|t_{\rm c.m.}(E_{\rm c.m.})|^2  \nonumber\\
&=&
\lambdabar^2 \sum_{\ell=0}^{\ell_{\rm max}} A_{\ell}(E_{\rm c.m.})
 P_\ell(\cos{\theta_{\rm c.m.}}),     
\label{eqn:e4}
\end{eqnarray}
where 
$t_{\rm c.m.}(E_{\rm c.m.})$ denotes the $K^- p \to K^+ \Xi^-$
$t$ matrix in the c.m.~frame, and 
$p_{f}$ ($p_{i}$) and $\omega_{f}$ ($\omega_{i}$) are a momentum 
and a reduced energy for $K^+ \Xi^-$ ($K^- p$) in the c.m.~frame, respectively; 
$\lambdabar$ is the de Broglie wavelength of $K^- p$, 
and $P_\ell(x)$ are Legendre polynomials.
Coefficient parameters  $A_{\ell}(E_{\rm c.m.})$ are expressed 
by a power series of $E_{\rm c.m.}$ so as to make a fit to 
their energy dependence.

Figure~\ref{fig:1} displays 
the angular distributions of $(d\sigma/d\Omega)^{\rm elem}_{\rm c.m.}$, 
together with the data at $E_{\rm c.m.}$ = 
1.948, 1.966, 2.069, 2.110, 2.134, 2.235, 2.276, 2.331, and 2.420 GeV
\cite{Trower68,Burgun68,Dauber69,Trippe67,London66}.
Here $\ell_{\rm max}=$ 6 is used.
Figure~\ref{fig:2} also displays the total (integrated) cross sections \cite{Flaminio79}
which are written as 
\begin{eqnarray}
\sigma_{\rm tot}(E_{\rm c.m.})
&=& \int d\Omega 
\Bigl( {d \sigma \over d \Omega} \Bigr)^{\rm elem}_{\rm c.m.} \nonumber\\
&=& 4\pi \lambdabar^2 A_0(E_{\rm c.m.}), 
\label{eqn:e4a}
\end{eqnarray}
as a function of $E_{\rm c.m.}$.
The parameters $A_{\ell}(E_{\rm c.m.})$ were determined for fits to 
242 data points for the angular distributions of $d\sigma/d\Omega$ 
in Fig.~\ref{fig:1}, 
together with 9 data points (open circles in Fig.~\ref{fig:2}) for the total cross sections 
of $\sigma_{\rm tot}$ which were simultaneously measured with the data of $d\sigma/d\Omega$. 
Thus the renormalized $\chi^2$ values account for  
$\chi^2/N=$ 1.48 with $N=$ 242 for $d\sigma/d\Omega$ 
and 
$\chi^2/N=$ 2.35 with $N=$ 56 (open and solid circles) for $\sigma_{\rm tot}$.

\begin{figure}[tb]
  \begin{center}
  \includegraphics[width=1.00\linewidth]{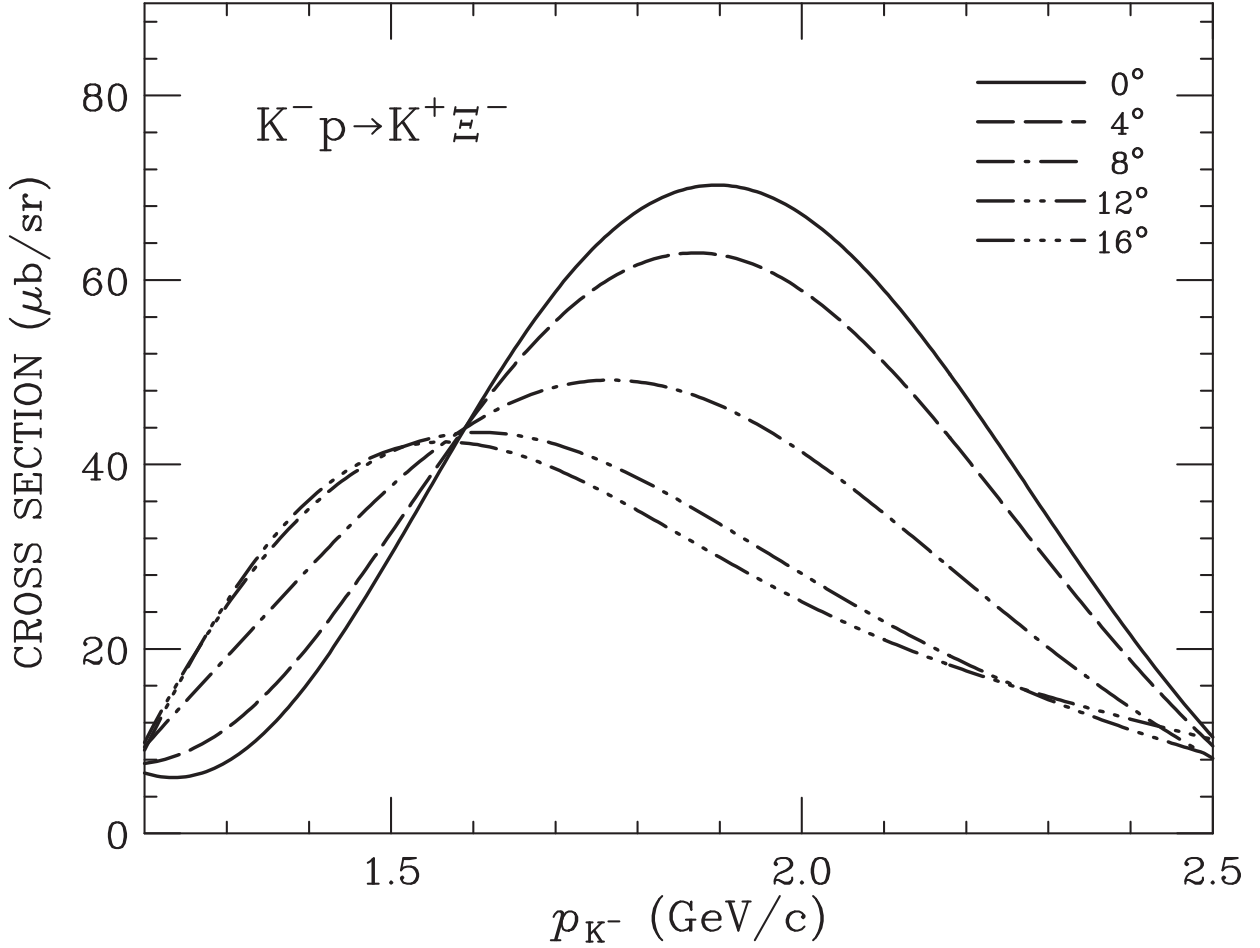}
  \end{center}
  \caption{\label{fig:3}
  Incident $K^-$ momentum dependence of the laboratory 
  differential cross sections for the $K^-p \to K^+ \Xi^-$ reaction, 
  $(d\sigma/d\Omega)^{\rm elem}$
  at $\theta_{\rm lab}=$ 0$^\circ$, 4$^\circ$, 8$^\circ$, 12$^\circ$, 
  and 16$^\circ$.
  }
\end{figure}

Figure~\ref{fig:3} shows the laboratory differential cross sections $(d\sigma/d\Omega)^{\rm elem}$
for the $K^-p \to K^+ \Xi^-$ reaction at $\theta_{\rm lab}=$ 0$^\circ$--16$^\circ$,
as a function of the incident $K^-$ momentum of $p_{K^-}$.
We find that the values of $(d\sigma/d\Omega)^{\rm elem}$ rather depend on $p_{K^-}$.
The peak of $(d\sigma/d\Omega)^{\rm elem}$ at $\theta_{\rm lab}=$ 0$^\circ$ 
is located at $p_{K^-}\sim$ 1.9 GeV/$c$, and its position is shifted downward
as $\theta_{\rm lab}$ increases; the peak of $(d\sigma/d\Omega)^{\rm elem}$ 
at $\theta_{\rm lab}=$ 16$^\circ$ stands at $p_{K^-}\sim$ 1.5 GeV/$c$.  
Thus this energy and angular dependence of $(d\sigma/d\Omega)^{\rm elem}$ 
may affect the shape and magnitude of the $\Xi^-$ production spectrum 
in the nuclear ($K^-$,~$K^+$) reaction at each $\theta_{\rm lab}$.

\section{Results and discussion}

\subsection{Optimal Fermi-averaged cross section}
\label{opt-app}

\begin{figure}[t]
  \begin{center}
  \includegraphics[width=0.80\linewidth]{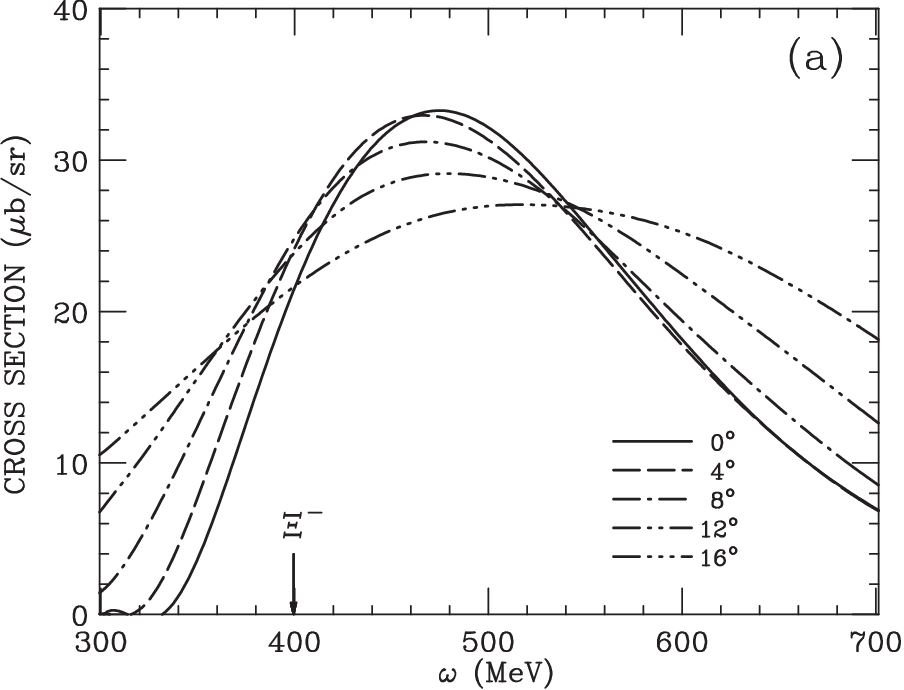}
  \includegraphics[width=0.80\linewidth]{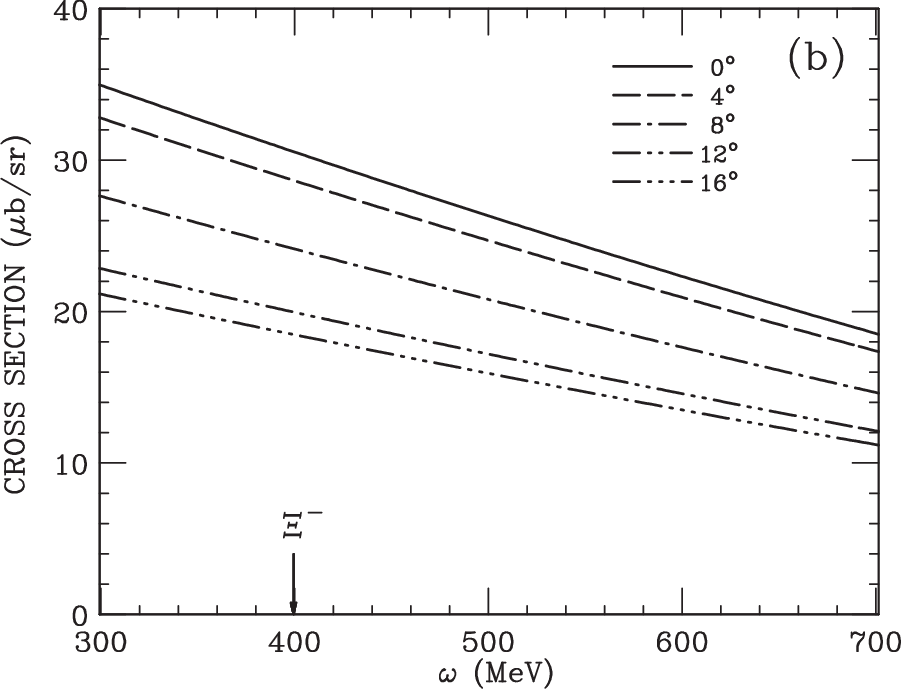}
  \end{center}
  \caption{\label{fig:4}
  Energy dependence of (a) the optimal Fermi-averaged cross section $(d \sigma/d \Omega)^{\rm opt}$
  and (b) the ordinary Fermi-averaged cross section of $\beta (d\sigma/d\Omega)^{\rm av}$
  with the kinematical factor $\beta$ of Eq.~(\ref{eqn:e2}) 
  for the $K^- p \to K^+ \Xi^-$ reaction on the $^{12}$C target
  at $p_{K^-}$=1.8 GeV/$c$ and $\theta_{\rm lab}=$ 0$^\circ$, 4$^\circ$, 
  8$^\circ$, 12$^\circ$, and 16$^\circ$, as a function of the energy transfer $\omega$ 
  . 
  The arrow shows the $\Xi^-$ emitted threshold.
  }
\end{figure}

To see the medium effects on the $K^-p \to K^+ \Xi^-$ reaction in nuclei, 
we evaluate the optimal Fermi-averaged cross section of $({d \sigma/d \Omega})^{\rm opt}$
in the laboratory frame, 
referring to the values of $(d\sigma/d\Omega)^{\rm elem}$ of Eq.~(\ref{eqn:e4}) 
in the optimal Fermi averaging procedure.
Figure~\ref{fig:4}(a) shows the calculated results of $({d \sigma/d \Omega})^{\rm opt}$ 
in the $K^-p \to K^+\Xi^-$ reaction at $p_{K^-}=$ 1.8 GeV/$c$ 
and $\theta_{\rm lab}=$ 0$^\circ$--16$^\circ$ 
in kinematics for a $^{12}$C target, as a function of the energy transfer $\omega$.
We find the strong energy dependence of $({d \sigma/d \Omega})^{\rm opt}$ for $\omega$, 
together with the angular dependence of them for $\theta_{\rm lab}$.
Note that the cross sections can be estimated in not only the $\Xi^-$ continuum region 
but also the $\Xi^-$ bound region due to the Fermi motion in the nuclear medium. 
The peak of $(d \sigma/d \Omega)^{\rm opt}$ 
is located at $\omega \simeq$ 480 MeV which corresponds to $E_{\Xi^-} \simeq$ 80 MeV 
with respect to the $\Xi^-$ emitted threshold when $\theta_{\rm lab}=$ 0$^\circ$--8$^\circ$, 
whereas the peak position moves upward as $\theta_{\rm lab}$ increases; 
when $\theta_{\rm lab}=$ 16$^\circ$, thus its position is located 
at $\omega \simeq$ 520 MeV ($E_{\Xi^-} \simeq$ 120 MeV). 
The behavior of $(d \sigma/d \Omega)^{\rm opt}$ can play a significant role 
in describing the shape and magnitude of the $\Xi^-$ spectrum 
in the nuclear ($K^-$,~$K^+$) reaction.
It was found that the energy and angular dependence of $({d \sigma/d \Omega})^{\rm opt}$ is 
significant to describe the behavior of the $\Lambda$ production spectra in the nuclear 
($\pi^+$,~$K^+$) reactions \cite{Harada04} 
and also the $\Sigma^-$ production ones in the nuclear ($\pi^-$,~$K^+$) reactions 
\cite{Harada05,Harada06,Harada18}.
Therefore, the optimal Fermi averaging procedure is expected to work well for 
describing the medium effects on the $K^-p \to K^+\Xi^-$ 
reaction in nuclei. 

In the standard DWIA in hypernuclear production theories \cite{Auerbach83,Dover80,Bando90}, 
we often find the ``ordinary'' Fermi averaging in which the Fermi-averaged differential 
cross section for the $K^-p \to K^+\Xi^-$ reaction may be given as
\begin{eqnarray}
\Bigl( {d \sigma \over d \Omega} \Bigr)^{\rm av}_{\theta_{\rm lab}}
&=& \int d{\bm p}_N \rho(p_N) \Bigl( {d \sigma \over d \Omega} \Bigr)^{\rm elem}.
\label{eqn:e7a}
\end{eqnarray}
In Fig.~\ref{fig:4}(b), we show the calculated cross sections of 
$\beta (d\sigma/d\Omega)^{\rm av}$ including the kinematical factor $\beta$ 
of Eq.~(\ref{eqn:e2}) at $p_{K^-}=$ 1.8 GeV/$c$ for $^{12}$C, in comparison with 
those of $({d \sigma/d \Omega})^{\rm opt}$. 
We confirm that the values of $\beta (d\sigma/d\Omega)^{\rm av}$ 
decrease monotonously due to the energy dependence of $\beta$ \cite{Tadokoro95}, 
as $\omega$ increases. Here we used, e.~g., the values of 
$(d\sigma/d\Omega)^{\rm av}=$ 51.9 $\mu$b/sr at $\theta_{\rm lab}=$ 4$^\circ$ 
and 43.1 $\mu$b/sr at $\theta_{\rm lab}=$ 8$^\circ$.
It is shown clearly that the values of $\beta (d\sigma/d\Omega)^{\rm av}$ 
are quite different from those of $({d \sigma/d \Omega})^{\rm opt}$ 
because the values of $(d\sigma/d\Omega)^{\rm av}$ depend on $p_{K^-}$ 
and $\theta_{\rm lab}$, not $\omega$. 

On the other hand, it is known that the impulse approximation (IA)
for nuclear reaction theories is improved to reduce the influence of 
the off-shell $t$ matrix caused by the Fermi motion for a nucleon in the nuclear target
by choosing an optimal momentum ${\bm p}_N^{\rm opt}$ for the nucleon \cite{Gurvitz86,Zhu87}. 
This is called the optimal momentum approximation \cite{Gurvitz86}
in which the use of the on-shell $t$ matrix may be valid 
because the leading order correction due to Fermi motion is minimized.
Several authors \cite{Zofka84,Sotona89} studied the effects of the Fermi motion 
in DWIA calculations for $\Lambda$ hypernuclear production, 
and proposed to use the optimal momentum ${\bm p}_N^{\rm opt}$ 
for the nucleon in the target nucleus.
This momentum may be given as ${\bm p}_N^{\rm opt}=(\eta-{1 \over 2}){\bm q}$ 
in the laboratory frame where $\eta$ is a factor determined at frozen point 
for Fermi motion \cite{Zhu87} instead of the Fermi averaging.  
However, their approach seems to be still insufficient to explain 
the influence of the energy and angular dependence of $(d\sigma/d\Omega)^{\rm elem}$ 
in the nuclear medium such as the $K^-p \to K^+\Xi^-$ reaction, 
in a quantitative comparison with the experimental data of 
the hypernuclear production~\cite{Zofka84}.
Consequently, we recognize that the optimal Fermi averaging procedure is 
a straightforward way by dealing with the Fermi-averaged amplitude 
for the elementary reaction in the optimal momentum approximation.

\subsection{Application to $\Xi^-$ QF production on $^{12}$C in a Fermi gas model}

To see the medium effects in the nuclear ($K^-$,~$K^+$) reaction, 
we demonstrate the $\Xi^-$ QF spectrum in the ($K^-$,~$K^+$) reaction on a nuclear target,  
using the calculated results of $({d \sigma/d \Omega})^{\rm opt}$ in Eq.~(\ref{eqn:e5}). 
We adopt a non-relativistic Fermi gas model \cite{deForest66}, 
in which each nucleon moves freely in the field of a uniform nuclear potential well $V$, 
for simplicity.
The QF spectrum in $\Lambda$ hypernuclear production via the ($K^-$,~$\pi^-$) reaction 
was first evaluated by Dalitz and Gal \cite{Dalitz76}. 
The $\Lambda$ QF spectrum for the associated ($\pi^+$,~$K^+$) reaction 
in terms of the high momentum transfer was also discussed by Dover et al.~\cite{Dover80}.
Following to Refs.~\cite{deForest66,Dover80}, 
the double differential cross section in the nuclear ($K^-$,~$K^+$) reaction in Eq.~(\ref{eqn:e1}) 
may be rewritten as 
\begin{eqnarray}
{{d}^2{\sigma} \over {d}E_{K^+}{d}\Omega_{K^+} }
&=& 
\Bigl( \displaystyle{d \sigma \over d \Omega} \Bigr)^{\rm opt}
R(\omega,{\bm q}).
\label{eqn:e8}
\end{eqnarray}
The response function $R(\omega,{\bm q})$ is defined as
\begin{eqnarray}
R(\omega,{\bm q})
&=&
Z_{\rm eff}{3 \over 4 \pi k_F^3}
\int {d{\bm p}_N}\theta(k_F-|{\bm p}_N|) \nonumber\\
&& \times 
\delta{\left(\bar{\omega}-{({\bm p}_N+{\bm q})^2 \over 2m_\Xi}
+{{\bm p_N}^2 \over 2m_N}\right)}, 
\label{eqn:e9}
\end{eqnarray}
where $k_{\rm F}$ is the Fermi momentum, 
${\bm p}_N$ is the momentum for the nucleon, 
and 
$\bar{\omega}=\omega-(m_\Xi-m_N)-(V_N-V_\Xi)$. 
In the eikonal approximation, 
the effective number of protons $Z_{\rm eff}$ may be 
approximated by 
\begin{eqnarray}
Z_{\rm eff}
&=& 
\int d{\bm r}\> \rho_{A}(r)|
\chi_{{\bm p}_{K^+}}^{(-)}({\bm r})|^2
|\chi_{{\bm p}_{K^-}}^{(+)}({\bm r})|^2 \nonumber\\
&\approx&
\int d{\bm b}\> T_{p}({\bm b})
\exp{[-{\bar{\sigma}}T({\bm b})]}, 
\label{eqn:e10}
\end{eqnarray}
where the nuclear thickness function is defined as
\begin{eqnarray}
T({\bm b})\equiv \int_{-\infty}^{\infty}\rho({\bm r})dz, \quad 
\int T({\bm b})d{\bm b}=A. 
\label{eqn:e11}
\end{eqnarray}
Here ${\bm b}$ is an impact-parameter coordinate in the plane perpendicular
to the direction of the momentum transfer ${\bm q}$; 
$T_p({\bm b})$ is a tickness function for the proton, 
thus $\int T_p({\bm b})d{\bm b}=Z$. 
The averaged total cross section for the $K^-N$ and $K^+N$ elastic scatterings 
is given as $\bar{\sigma}=\tfrac{1}{2}(\sigma_{K^-N}+\sigma_{K^+N})$.
Note that $Z_{\rm eff}$ reduces to the proton number $Z$
when the limit of no distortion ($\bar{\sigma} \to 0$).

Considering $q > k_F$ because $q \simeq$ 390$-$800 MeV/$c$ 
($\theta_{\rm lab}=$ 0$^\circ$--16$^\circ$) in the nuclear 
($K^-$,~$K^+$) reaction, 
the response function in Eq.~(\ref{eqn:e9}) is easily obtained as
\begin{eqnarray}
&&R(\omega,{\bm q})=
Z_{\rm eff}{3  \over 4 \pi}\left(m_\Xi \over k_F^2\right) 
{\pi \over \sqrt{Q^2-4\alpha^2(\nu-Q^2/2)}} \nonumber\\
&&\times \left\{1-{1 \over 4\alpha^4} 
\left(Q-\sqrt{Q^2-4\alpha^2(\nu-Q^2/2)}\right)^2\right\} \nonumber\\
\label{eqn:e15}
\end{eqnarray}
for $Q > 2\alpha^2$, where the dimensionless valuables of $Q=q/k_F$, 
$\alpha^2= (m_\Xi-m_N)/2m_N$, $\nu= m_\Xi\bar{\omega}/k_F^2$; 
we confirm that for small $Q$, 
$R(\omega,{\bm q})$ is proportional to $1/Q\{1-(\nu/Q-Q/2)^2\}$, 
a well-known parabolic function \cite{deForest66} 
as well as for $\alpha^2 \to 0$ ($m_\Xi/m_N \to 1$).
The peak of this response occurs at $\nu= Q^2/2 + \alpha^2$ or 
\begin{eqnarray}
{\omega}_{\rm peak}
&=& m_\Xi-m_N+(V_N-V_\Xi)   \nonumber\\
&+& {m_\Xi-m_N \over m_\Xi}{k_F^2 \over 2m_N}+{q^2 \over 2 m_\Xi},
\end{eqnarray}
and its width $2\Delta= 2 k_F q/m_\Xi$, 
whereas this position may be moderated in the $\Xi^-$ spectrum of Eq.~(\ref{eqn:e8}) 
due to the energy dependence of $(d\sigma/d\Omega)^{\rm opt}$ or $\beta(d\sigma/d\Omega)^{\rm av}$.

\begin{figure}[t]
  \begin{center}
  \includegraphics[width=0.80\linewidth]{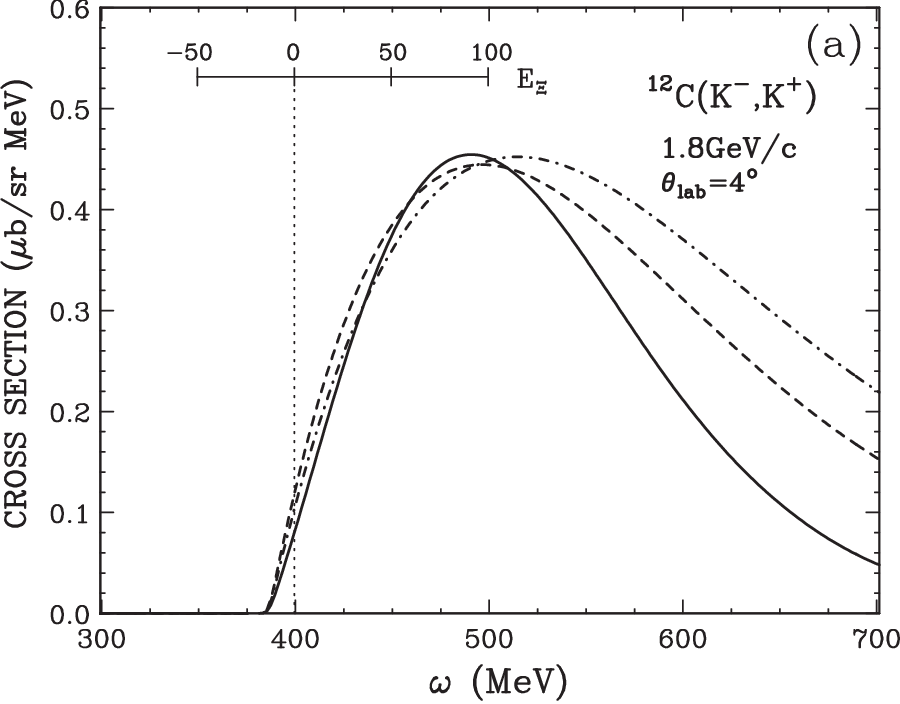}
  \includegraphics[width=0.80\linewidth]{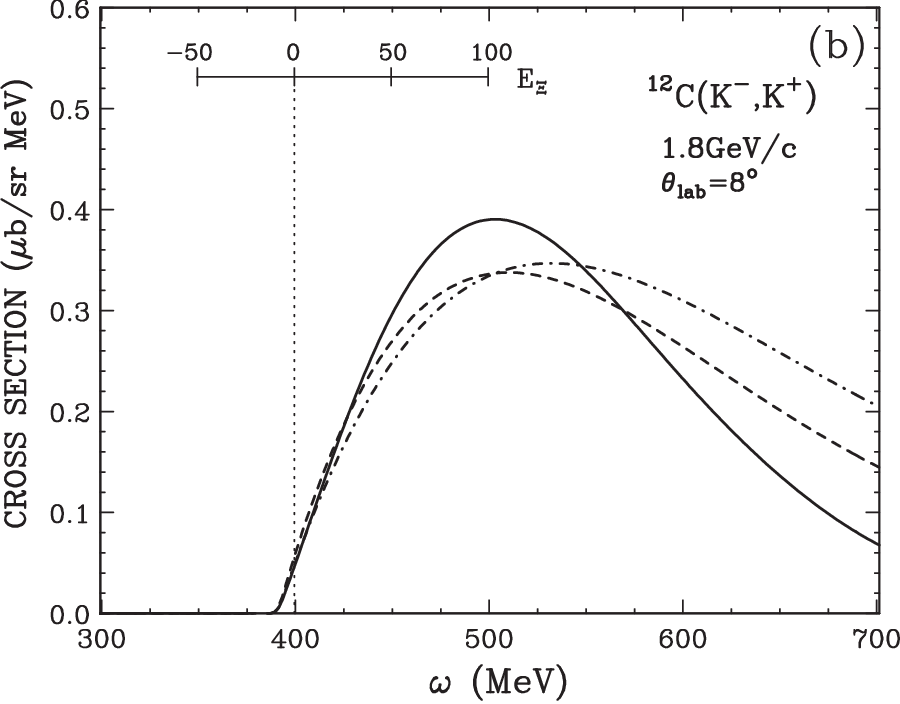}
  \end{center}
  \caption{\label{fig:5}
  The $\Xi^-$ QF production spectra in the Fermi gas model 
  in the $^{12}$C($K^-$,~$K^+$) reaction at $p_{K^-}$= 1.8 GeV/$c$,  
  (a) $\theta_{\rm lab}=$ 4$^\circ$ and (b) 8$^\circ$, 
  as a function of the energy transfer $\omega$. 
  Solid, dashed, and dot-dashed curves denote the calculated results of 
  the spectra with $(d\sigma/d\Omega)^{\rm opt}$, $\beta (d\sigma/d\Omega)^{\rm av}$, 
  and the constant value of $\beta (d\sigma/d\Omega)^{\rm av}=$ 31.1 (25.9) $\mu$b/sr 
  at $\theta_{lab}=$ 4$^\circ$ (8$^\circ$), respectively. 
  The spectra are folded with a detector resolution of 5 MeV FWHM.
  }
\end{figure}

Now let us estimate the $\Xi^-$ QF spectrum in the ($K^-$,~$K^+$) reaction on 
the $^{12}$C target at $p_{K^-}=$ 1.8 GeV/$c$.
For distortion, we use $\bar{\sigma}=$ 24.2 mb as the parameter, 
where $\sigma_{K^-}$= 28.9 mb for $K^- N$ scattering 
and $\sigma_{K^+}$= 19.4 mb for $K^+ N$ one \cite{Harada05,Harada06}.
When we use the nuclear MHO density for $^{12}$C which is written as 
\begin{eqnarray}
\rho({\bm r})= \rho_0\{1+a({\bm r}/d)^2\}\exp{\{-({\bm r}/d)^2\}}
\label{eqn:e13}
\end{eqnarray}
with $a=$ 2.234 and $d=$ 1.516 fm \cite{Friedma94}, 
we find $Z_{\rm eff}=$ 2.09, evaluating Eq.(\ref{eqn:e10}) numerically 
by the eikonal-wave integral \cite{Dover80,Dover83}.
For $p_{K^-}=$ 1.8 GeV/$c$ and $\theta_{\rm lab}=$ 4$^{\circ}$, 
thus, we estimate $\bar{\omega}_{\rm peak} \simeq$ 542 MeV and $2\Delta \simeq$ 222 MeV consistently, 
using $q\simeq$ 542 MeV/$c$ at $\omega=$ 537 MeV for $^{12}$C.  Here we took masses of 
$m_N=$ 938.27 MeV for the proton and $m_\Xi=$ 1321.71 MeV for the $\Xi^-$ hyperon, 
the attractive well depths of $V_N=$ 50 MeV and $V_\Xi=$ 14 MeV, 
and the Fermi momentum of $k_F=$ 270 MeV/$c$ in the nucleus.
We also consider the recoil correction in the spectrum, 
replacing $q$ by $q_{\rm eff}=(M_C/M_A)q$ 
where $M_A$ and $M_C$ denote the masses of 
the $^{12}$C target and the $^{11}$B core nuclei, respectively \cite{Harada19}. 
 
In Fig.~\ref{fig:5}, we show the calculated $\Xi^-$ QF spectra 
of Eq.~(\ref{eqn:e8}) 
in the $^{12}$C($K^-$,~$K^+$) reaction at $p_{K^-}$= 1.8 GeV/$c$, 
$\theta_{\rm lab}=$ 4$^\circ$ and 8$^\circ$.
The calculated spectrum at $\theta_{\rm lab}=$ 4$^\circ$ has 
a QF peak at $\omega \simeq$ 480 MeV  which corresponds to about 80 MeV 
above the $\Xi^-$ emitted threshold, and 
its width of $2\Delta \simeq$ 150 MeV which is extremely narrower than that 
of $2\Delta \simeq$ 200 MeV for the spectrum with $\beta (d\sigma/d\Omega)^{\rm av}$.
For $\theta_{\rm lab}=$ 8$^\circ$, the QF peak position slightly shifts upward 
($\omega \simeq$ 500 MeV) and its width becomes $2\Delta \simeq$ 200 MeV.  
We find that the $\omega$ dependence of 
$(d\sigma/d\Omega)^{\rm opt}$ acts on the shape and magnitude 
of the QF spectrum remarkably, and it makes its width narrower, 
in comparison with that of $\beta (d\sigma/d\Omega)^{\rm av}$ 
as well as that in the case of 
$\beta (d\sigma/d\Omega)^{\rm av}=$ const.~which is proportional to $R(\omega, {\bm q})$.
The magnitude of the spectrum seems to be rather as large as those with 
$\beta (d\sigma/d\Omega)^{\rm av}$ in the forward angles 
($\theta_{\rm lab}\leq$ 4$^\circ$), whereas we find a considerable 
difference among the shapes of these spectra near the $\Xi^-$ emitted threshold. 
Consequently, we realize that the optimal Fermi-averaged amplitudes of 
$\overline{f}_{K^-p \to K^+\Xi^-}$ in Eq.~(\ref{eqn:e3}) provide 
to moderate directly the shape and magnitude of the spectrum including 
the $\Xi^-$ QF region with a wide energy range.
Thus it is required to extract information concerning the $\Xi$-nucleus 
potential carefully from the data of the experimental spectrum.

Moreover, it should be mentioned that 
there still remain some uncertainties about the values of $(d\sigma/d\Omega)^{\rm elem}$ 
due to the limit of the available data of the $K^-p \to K^+\Xi^-$ reaction.
Thus if we use the different parameters of $A_\ell(E_{\rm c.m.})$ in Eq.~(\ref{eqn:e4a}) 
which simulate the values of $\sigma_{\rm tot}$ predicted in Ref.~\cite{Jackson15}, 
we find that the shape of the calculated values of $(d\sigma/d\Omega)^{\rm opt}$ 
is modified and the magnitude is reduced by about 20\% 
in the region of $\omega \simeq$ 400--500 MeV above the $\Xi^-$ emitted threshold.
This implies the ability of judging the validity of $(d\sigma/d\Omega)^{\rm elem}$
if our approach satisfies completely a description of the $\Xi^-$ spectrum 
in the ($K^-$,~$K^+$) reaction. 
An attempt to such studies may need a more quantitative observation 
of the differential cross sections 
in the elementary $p$($K^-$,~$K^+$)$\Xi^-$ reaction and the nuclear ($K^-$,~$K^+$) 
reaction experimentally. 

\subsection{Applicability of the optimal Fermi-averaging}

We study the applicability of the optimal Fermi-averaging procedure to the inclusive ($K^-$,~$K^+$) reactions
by high momentum $K^-$ beams 
because the contribution of two-step processes may grow in the inclusive reactions 
on heavier nuclear targets \cite{Iijima92,Nara97}. 
We recognize that the optimal Fermi-averaged $t$ matrix 
provides the benefit of the optimal momentum approximation \cite{Gurvitz86}, 
as already mentioned in Sect.~\ref{opt-app}.
Considering the general relation between the exact scattering operator 
$\tau$ and an approximate one $t_a$, which is written by 
\begin{eqnarray}
\tau &=& t_{a} +t_{a} G_a h G_a t_a \nonumber\\
&& + t_a G_a h(G_a + G_a t_a G_a )h G_a t_a + \cdots 
\end{eqnarray}
in terms of the expansion series with $h \equiv G_a^{-1}-G^{-1}$ 
where $G$ and $G_a$ are exact and approximate Green's functions, respectively, 
we confirm that the contribution from the first-order correction term of $t_{a} G_a h G_a t_a$ 
vanishes in the optimal momentum approximation \cite{Gurvitz86}. 
As far as the $\Xi^-$ production is concerned in the inclusive ($K^-$, $K^+$) reactions, 
this correction may correspond to the two-step processes of $K^-N \to K^-N$ followed by $K^-p \to K^+\Xi^-$, 
or $K^-p \to K^+\Xi^-$ followed by $K^+N \to K^+N$, involving rescattering effects in nuclear medium. 
It implies that the optimal Fermi-averaged $t$ matrix ${t}^{\rm opt}_{\bar{K}N,K\Xi}$
 inevitably includes medium effects due to these two-step processes. 
Therefore, we expect that the optimal Fermi-averaging procedure works well within the impulse approximation 
for the $\Xi^-$ production on light nuclei, and also on heavier nuclei in which 
strong absorption and distortion effects must be taken into account.

On the other hand, if we consider strangeness productions with various 
hyperons ($Y=$ $\Xi$, $\Xi^*$, $\Lambda$, $\Sigma$, $\cdots$) in the inclusive ($K^-$,~$K^+$) reactions, 
it is necessary to deal with additional optimal Fermi-averaged ${t}$ matrices 
for the corresponding reactions, e.g, $\bar{K}N \to \bar{K}N$, $KN \to KN$, $\bar{K}N \to YM$ ($Y^*M$) 
followed by $MN \to YK$ ($Y^*K^*$), 
because the contributions of the two-step processes with intermediate mesons 
($M=$ $\pi$, $K$, $\bar{K}$, $\phi$, $a_0$, and $f_0$) and hyperon resonances are very important \cite{Nara97}. 
We believe that our procedure can be extended to the additional two-step processes 
for the strangeness productions in the inclusive ($K^-$,~$K^+$) reactions.

\section{Summary and Conclusion}
\label{summary}

We have studied theoretically the medium effects on 
the $\Xi^-$ production via the $K^-p\to K^+\Xi^-$ process in the nuclear ($K^-$,~$K^+$) reaction, 
using the optimal Fermi-averaging procedure.
The calculated optimal Fermi-averaged amplitudes of $\overline{f}_{K^-p \to K^+\Xi^-}$ 
for our DWIA calculations provide the strong energy and angular dependence, 
leading to the fact that the shape and magnitude of 
the $\Xi^-$ production spectrum is influenced in the ($K^-$,~$K^+$) reaction 
on a nuclear target. 
We have also demonstrated the $\Xi^-$ QF spectrum in
the $^{12}$C($K^-$,~$K^+$) reaction at $p_{K^-}=$ 1.8 GeV/$c$ 
within the Fermi gas model.

In conclusion, we show 
the strong energy and angular dependence of the in-medium $K^-p\to K^+\Xi^-$ production cross section, 
which is important to describe the shape and magnitude of the $\Xi^-$ production spectrum 
in the ($K^-$,~$K^+$) reaction on the nuclear target. 
This result may be a basis for study extracting the properties 
of the $\Xi$-nucleus potential from the experimental data. 
The detailed investigations are required for the analysis of the 
$^{12}$C($K^-$,~$K^+$) reaction at 1.8 GeV/$c$ at the J-PARC E05 experiment \cite{Nagae18}, 
and also for the extension to the two-step processes in the inclusive ($K^-$,~$K^+$) reactions. 
These investigations are in progress. 
\\

\begin{acknowledgments}
The authors would like to thank Professor~T.~Fukuda, 
Professor~T.~Nagae, and Professor~Y.~Akaishi for many valuable discussions.
This work was supported by Grants-in-Aid for 
Scientific Research (KAKENHI) from the Japan Society for 
the Promotion of Science (Grant No.~JP20K03954).
\end{acknowledgments}



\clearpage

\end{document}